\documentclass{aastex61}
\usepackage[ampersand]{easylist}
\usepackage{xcolor}
\usepackage{soul}

\begin{document}
\title{A Monte Carlo Method for Evaluating Empirical Gyrochronology Models and its Application to Wide Binary Benchmarks} 

\correspondingauthor{Tomomi Otani}
\email{otanit@erau.edu}

\author[0000-0002-1691-8479]{Tomomi Otani}
\affil{Department of Physical Sciences, Embry-Riddle Aeronautical University, 1 Aerospace Blvd, Daytona Beach, FL 32114, United States}

\author[0000-0002-5775-2866]{Ted von Hippel}
\affil{Department of Physical Sciences, Embry-Riddle Aeronautical University, 1 Aerospace Blvd, Daytona Beach, FL 32114, United States}

\author[0000-0002-1988-143X]{Derek Buzasi}
\affiliation{Department of Chemistry and Physics, Florida Gulf Coast University, 10501 FGCU Blvd S, Fort Myers, FL 33965, United States}

\author[0000-0002-8541-9921]{T. D. Oswalt}
\affiliation{Department of Physical Sciences, Embry-Riddle Aeronautical University, 1 Aerospace Blvd, Daytona Beach, FL 32114, United States}

\author{Alexander Stone-Martinez}
\affiliation{Department of Physical Sciences, Embry-Riddle Aeronautical University, 1 Aerospace Blvd, Daytona Beach, FL 32114, United States}
\affiliation{Department of Astronomy, New Mexico State University, 1780 E University Ave, Las Cruces, NM 88003, United States}

\author{Patrice Majewski}
\affiliation{Department of Physical Sciences, Embry-Riddle Aeronautical University, 1 Aerospace Blvd, Daytona Beach, FL 32114, United States}
\affiliation{Whiting School \& Engineering, Johns Hopkins University, 3400 North Charles Street, Baltimore, MD 21218, United States}

\accepted{Mar 21, 2022}
\submitjournal{ApJ}
\begin{abstract}

Accurate stellar ages are essential for our understanding of the star formation history of the Milky Way, Galactic chemical evolution, and to constrain exoplanet formation models. Gyrochronology, a relationship between stellar rotation and age, appears to offer a reliable age indicator for main sequence (MS) stars over the mass range of approximately 0.6 to 1.3 $M_\sun$.  Those stars lose their angular momentum due to magnetic braking and as a result their rotation speeds decreases with age.  Although current gyrochronology relations are fairly well tested for young MS stars with masses greater than 1 $M_\sun$, primarily in young open clusters, insufficient tests exists for older and lower mass MS stars.  Binary stars offer the potential to expand and fill in the range of ages and metallicity over which gyrochronology can be empirically tested. In this paper, we demonstrate a Monte Carlo approach to evaluate gyrochronology models using binary stars.  As examples, we used five previously published wide binary pairs. We also demonstrate a Monte Carlo approach to assess the precision and accuracy of ages derived from each gyrochronology model. For the traditional Skumanich models, the age uncertainties are $\sigma_{age}$/$age$ = 15-20\% for stars with $B-V$ = 0.65, and $\sigma_{age}$/$age$ = 5-10\% for stars with $B-V$ = 1.5 and rotation period P $\leq$ 20 days. 

\end{abstract}

\section{Introduction} \label{sec:intro}
Ages are among the most difficult stellar properties to determine, yet they provide key constraints to problems ranging from the formation and habitability of exoplanets to the Galaxy’s chemical evolution and star formation history.  For decades it has been known that main sequence (MS) stars of spectral type F through early M have outer convection zones with strong magnetic fields and that they lose angular momentum due to magnetic braking \citep{Schatzman1962, Weber1967, Skumanich1972, Kawaler1988, Pallavicini1981, Soderblom1985, Pinsonneault1989}.  This magnetic braking among low mass MS stars was observed by \citet{Skumanich1972}, who found that the rotation period was proportional to the square root of stellar age.  This is now referred to as the Skumanich law and this phenomenon of a declining rotation rate with time is referred to as the slow rotator sequence.  Subsequently, both empirical/semi-empirical \citep{Barnes2001, Barnes2003, Barnes2007, Barnes2010, Barnes_Kim2010, Brown2014, Angus2015, Gondoin2017, Angus2019} and theoretical \citep{Kawaler1988, Kawaler1989, Pinsonneault1989, vanSaders2013, Matt2015, vanSaders2016, Spada2020} frameworks for measuring stellar ages from rotation periods, now called gyrochronology, have been developed.  Numerous researchers have made the case that gyrochronology could be developed into one of the most precise stellar chronometric techniques for $\sim$0.6–1.0 M$_\odot$ field stars \citep{Barnes2003, Soderblom2010, Epstein2014, vanSaders2016}. 


When a star arrives on the zero age main sequence, its rotational period is typically between one and ten days \citep{Irwin2009, Gallet2013}.  Rotation periods begin to follow the Skumanich law when stars are 500 to 700 Myr old, and by the age of 1 Gyr, all stars with masses between 0.6 and 1.3 $M_\odot$ will join this sequence. Although open clusters confirm that the gyrochronology paradigm is valid for the Sun and lower mass stars, there is ongoing concern about how age precision is affected by such factors as the initial range of rotation rates \citep{Barnes2010, Matt2012}, the choice of spin-down model \citep{Aigrain2015}, changing magnetic morphology \citep{Buzasi1997, Garraffo2018}, and how mass loss rates may change for stars older than the Sun \citep{vanSaders2016, Metcalfe2019}. In addition, evidence is accumulating that the rate of rotational spin-down may temporarily stall and does not follow the traditional Skumanich law \citep{Meibom2011b, Meibom2015, McQuillan2014, Rebull2017, Curtis2019}. 
\citet{Curtis2019} found that both early and late K stellar rotation rates were the same in the open cluster NGC 6811 ($\sim$1 Gyr).  \citet{Curtis2019} and \citet{Douglas2019} compared the open clusters NGC 6811 and Praesepe and note that despite their age differences both low mass star sequences below 0.8 $M_\odot$ seem to overlap, and furthermore that low mass stars in NGC 6811 may not have spun down for the past 300 Myr. \citet{Angus2020} and \citet {Spada2020} suggests that the angular momentum transport from the core to the envelope of stars causes this extended period over which rotation rates do not decrease.
Additionally, evidence is accumulating that field stars older than $\sim$1 Gyr rotate more rapidly than predicted by the Skumanich law \citep{Angus2015, vanSaders2016, vanSaders2018, Metcalfe2019}. \citet{Curtis2020} studied the dispersed star cluster Ruprecht 147 ($\sim$2.7 Gyr) and found that its gyrochrone is flat compared to other clusters with younger ages. Also, \citet{Agueros2018} suggests that the slow rotator sequence of the open cluster NGC 752 (1.3 Gyr) departs from the classical Skumanich law. Both \citet{vanSaders2016} and \citet{vanSaders2018} suggest that magnetic braking would stop when stars become sufficiently old that their Rossby number ($R_O$), the ratio of rotation period to convective turnover time, reaches approximately two. 
Therefore, while the traditional Skumanich law may work well for younger F and G stars, it may not apply to cooler and older stars. For these reasons, the field of gyrochronology requires both a wider range of empirical tests and further theoretical development.

Empirical tests based on ages derived from wide non-interacting binary stars offer several advantages.  First, components of each pair are coeval and should exhibit the same rotation age.  Many thousands of nearby pairs are potentially useful for testing gyrochronology.  Among the 1.3 million wide binaries found in the $Gaia$ EDR3 by \citet{Elbadry2021}, about 22,000 are within 100 pc. The sheer number of stars in open clusters offers substantial advantages, yet due to their distances, low-mass members of many clusters such as NGC 6811, NGC 752, NGC 6819, Ruprecht 147, and M67, are too faint to readily observe.  Additionally, binary pairs span a well-sampled range of ages, extending beyond the time most open clusters have lost most low mass members and they  span a wide range of metallicities.   Finally, unlike asteroseismology, which with current technology is limited to bright stars with a relatively narrow range of masses, wide pairs contain stars that span the entire range of masses over which gyrochronology is believed to apply. 

Gyrochronology requires measured rotation periods.  The $Kepler$ missions, with their continuous monitoring of selected star fields, have obtained data useful for deriving the rotation periods for thousands of stars.  Unfortunately, the vast majority of $Kepler$ and $K2$ stars do not have $B-V$ color measurements, the most common proxy for mass upon which most current gyrochronology models are based.  For about 100 wide pairs in the original $Kepler$ field, \citet{Janes2017} constructed an empirical $B-V$ vs. $g-K$ relation. A more recent search for wide pairs in the Kepler field was conducted by \citet{Godoy-Rivera2018}.  This study used $Gaia$ DR2 astrometry, parallaxes, radial velocities and metallicities to identify candidate pairs.

In this paper we develop a Monte Carlo strategy for propagating uncertainties in the observed stellar parameters and theoretical model coefficients in order to quantify the resulting age uncertainties, then propose a model evaluation method using wide binaries, under the assumption that binary components are coeval. As an application of this strategy, we chose six gyrochronology models that are based on rotation periods and colors \citep{Barnes2007, MH2008, Meibom2011a, Angus2015, Angus2019, Spada2020}.  (Hereafter we refer to these models as BA07, MH08, M11, A15, A19, and SL20, respectively). 

Where age inconsistencies between binary components arise, they may be caused by a variety of observational or theoretical reasons:
\begin{easylist}[itemize]
& one or both members of the pair suffer from exceptionally large errors in color or rotation period,
& one or both members of the pair lie outside the region of parameter space in which gyrochronology is applicable,
& undetected tertiary components modified the evolutionary history, mass, or rotation period of the observed star,
& blending of the target star with another star along the line of sight, 
& the presence of multiple stellar spots or groups yielding an incorrect rotation period, 
& spots measured at different latitudes and/or differential rotation, 
& pulsation periods masquerading as rotation periods, and 
& problems with the gyrochronology models such as  uncertainties, fitting parameters inaccuracies, and missing physics in gyrochronology model.
\end{easylist}
A precise determination of both consistencies and inconsistencies among stars with gyrochronological ages will help us investigate these issues in order to improve both the observational techniques and the models.

The models that were used in this manuscript is shown in Section~\ref{sec:models}. The uncertainty sensitivity of the five empirical models (B07, MH08, M11, A15, and A19) are discussed in Section~\ref{sec:one_ex}. The model evaluation method is discussed in Section~\ref{sec:method}. Application of the method using the six gyrochronology models and an evaluation of the age precision of those models are discussed in Section~\ref{sec:result}. The conclusions drawn from these numerical experiments are described in Section~\ref{Conclusions}.

\section{Gyrochronology Models}
\label{sec:models}

\citet{Barnes2003} first proposed an empirical relationship between the rotation period, color, and age of the form: 

\begin{equation}
P = \sqrt{t} \: f(B-V)
\label{eq:1}
\end{equation}

where

\begin{equation}
f(B-V) = \sqrt{B-V - 0.5} - 0.15(B-V - 0.5)
\label{eq:2}
\end{equation}

Here, $P$ is the rotation period in days, $t$ is the age of the star in Myr and, $B-V$ is the color index. Later, \citet{Barnes2007} updated the relationship as:

\begin{equation}
P = A^n \times a~(B-V - c)^b
\label{eq:3}
\end{equation}
where $A$ is the age of the star in Myr, and $a$, $b$, $c$, and $n$ are dimensionless free parameters. The four empirical models that are discussed in this section (B07, MH08, M11, and A15) use this simple relationship. The parameters used by each of these empirical models are shown in Table~\ref{tab:coefficients}. To estimate the age of a star, $P$ and $B-V$ must be obtained from observations. Nearly continuous photometry from space telescopes ($Kepler$, $K2$, and $TESS$) have provided rotation periods for thousands of stars. The majority of $B-V$ color indices for these stars have come from ground-based photometry. 


\begin{deluxetable}{rrrrr} [ht!]
\tablecaption{Coefficients for four models.}\label{tab:coefficients}
\tablenum{1}
\tablehead{\colhead{coefficients} & \colhead{B07} & \colhead{MH08}& \colhead{M11}  & \colhead{A15} \\ } 
\startdata
a                & 0.7725 & 0.407 & 0.700  & 0.40 \\
$+$ $\sigma_{a}$ & 0.011  & 0.021 & 0.013  & 0.30 \\
$-$ $\sigma_{a}$ & 0.011  & 0.021 & 0.013 & 0.05 \\ \hline
b                & 0.601  & 0.325 & 0.553 & 0.31 \\
$+$ $\sigma_{b}$ & 0.024  & 0.024 & 0.052 & 0.05 \\
$-$ $\sigma_{b}$ & 0.024  & 0.024 & 0.052 & 0.02 \\ \hline
c                & 0.4    & 0.495 & 0.472 & 0.45 \\
$+$ $\sigma_{c}$ & N/A    & 0.01  & 0.027  & N/A \\
$-$ $\sigma_{c}$ & N/A    & 0.01  & 0.027  & N/A \\ \hline
n                & 0.5189 & 0.566 & 0.52 & 0.55 \\
$+$ $\sigma_{n}$ & 0.007  & 0.008 & N/A & 0.02 \\
$-$ $\sigma_{n}$ & 0.007  & 0.008 & N/A & 0.09 \\
\enddata
\end{deluxetable}

\begin{deluxetable}{rr} [ht!]
\tablecaption{Model coefficients for A19.}\label{tab:coefficients2}
\tablenum{2}
\tablehead{\colhead{\hspace{1.75cm}coefficients} & \colhead{\hspace{1.75cm}A19}\\ } 
\startdata
$c_A$                & 0.65 \\
$\sigma_{c_A}$ & 0.05  \\\hline
$c_0$                & -4.7 \\
$\sigma_{c_0}$ & 0.5  \\\hline
$c_1$                & 0.72 \\
$\sigma_{c_1}$ & 0.05  \\\hline
$c_2$                & -4.9 \\
$\sigma_{c_2}$ & 0.2  \\\hline
$c_3$                & 29 \\
$\sigma_{c_3}$ & 2  \\\hline
$c_4$                & -38  \\
$\sigma_{c_4}$ & 4  \\\hline
\enddata
\end{deluxetable}

Solving Equation~\ref{eq:3} for stellar age, adding a subscript to age, $A_g$, to acknowledge that this is not necessarily the true age of the star, but rather the age provided via a gyrochronology relation, we have 

\begin{equation}
D(A_g) = \left\{ \frac{G(P)}{D(a) \times \left[ G(B-V)-D(c) \right] ^{D(b)}} \right\} ^{\frac{1}{D(n)}},
\label{eq:4}
\end{equation}
where $G(B-V)$ and $G(P)$ indicate that these observable parameters have uncertainties characterized by Gaussian distributions using the observed values and one-sigma uncertainties and $D(a,b,c,n)$ indicates that the model parameter uncertainty distributions may not be Gaussian.  


The Skumanich models do not always provide reliable results, particularly for low mass main sequence stars \citep{Meibom2011b,Meibom2015,McQuillan2014,Garcia2014, Douglas2017, Rebull2017, Curtis2019}. This may be due to angular momentum transport from stellar interiors to their surfaces \citep{Angus2020, Spada2020}.  Although stellar surfaces continuously lose angular momentum via magnetic wind braking, this additional angular momentum from the interior compensates for that loss, and it thus $appears$ that these star do not lose angular momentum during this age range. In this paper, we use two models that include this angular momentum transport (A19 and SL20). 

The A19 model improves the age precision for F, G, and K dwarfs, including stars that do not always follow the Skumanich-style slow rotator sequence. A19 also includes subgiant stars that were not included in previous gyrochronology models. Because this model uses $Gaia$ $G_{BP}-G_{RP}$ colors, which are available for more than a billion stars \citep{Gaia2018}, it is easier to find accurate and consistent color information than for those models that rely on $B-V$ photometry.  The A19 model was derived by combining both isochrones and empirical gyrochronology models using data from Praesepe ($\sim$700 Myr) and the Sun. In order to incorporate A19 in a consistent manner with the other models, we only used the gyrochronology component of the full A19 model so only $B-V$ and P are required to estimate ages.  A19 warns that the gyrochronology models only works for late F, G, K and early M dwarfs (0.56 $<$ $G_{BP}-G_{RP}$ $<$ 2.7) with Rossby numbers less than 2. The relationship between rotation periods and age of the stars is:

\begin{equation}
log_{10}(P) = c_A log_{10}(t) + \sum_{0}^{4} c_n[log_{10}(G_{BP}-G_{RP})]^n
\label{eq:5}
\end{equation}

\noindent where P is rotation period in days, t is age of the star in years, $G_{BP}$-$G_{RP}$ is the $Gaia$ color, and each coefficient is shown in Table~\ref{tab:coefficients2}.  Ages from their model can be obtained from an open-source $Python$ package, $stargate$, which we used (see details in A19).

SL20, updating their previous work \citep{Lanzafame2015}, construct gyrochrones using the 0.4-1.3 $M_\odot$ star data from Praesepe ($\sim$700 Myr) and NGC6811 ($\sim$1 Gyr).  Their model includes the reduced spin-down observed in NGC 6811 due to the redistribution of angular momentum from the stellar interior. Therefore it is built not only with the magnetic wind braking law, but also with the semi-empirical mass dependence of the rotational coupling time scale \citep{Lanzafame2015, Somers2016}. This model agrees with the stalled surface spin-down which was observed among low mass stars in Praesepe and NGC6811.

\section{Uncertainty Sensitivity for Individual Stars} \label{sec:one_ex}

To propagate errors, we employ a Monte Carlo approach, allowing us to examine the age precision achievable with these empirical models throughout the color-period diagram.  In this approach, we drew 500 values of color ($B-V$) and rotation period ($P$) at grid points covering $B-V$ = 0.6-2.3 mag (or $G_{BP}$ - $G_{RP}$ = 0.5-2.7) and $P$ = 0-60 days.  For each color ($B-V$ or $G_{BP}-G_{RP}$) and rotation period ($P$) combination, we obtained 10000 random normal distribution samples of color ($B-V$), rotation period ($P$), and gyrochronology model coefficients using their mean values and the standard deviations. For A15, two Gaussian distributions were created using both the positive and negative uncertainties, to create this non-Gaussian distribution. We did not include the effects of correlations between the parameters a and n for A15. We also did not include the A19 and SL20 models in this uncertainty analysis because these models did not include their uncertainties.  We calculated ages for each combination using the method discussed in Section~\ref{sec:models}, then made distribution of the ages using the 10000 simulated data points propagated through the appropriate model.  From these  distributions, we obtained median ages ($age$) and standard deviations ($\sigma_{+}$ and $-\sigma_{-}$).  The mean of the absolute values of $\sigma_{+}$ and $-\sigma_{-}$ are used to estimate the age uncertainties $\frac{\sigma_{age}}{age}$.



\begin{figure}[ht!]
\begin{center}
\includegraphics[width=\textwidth]{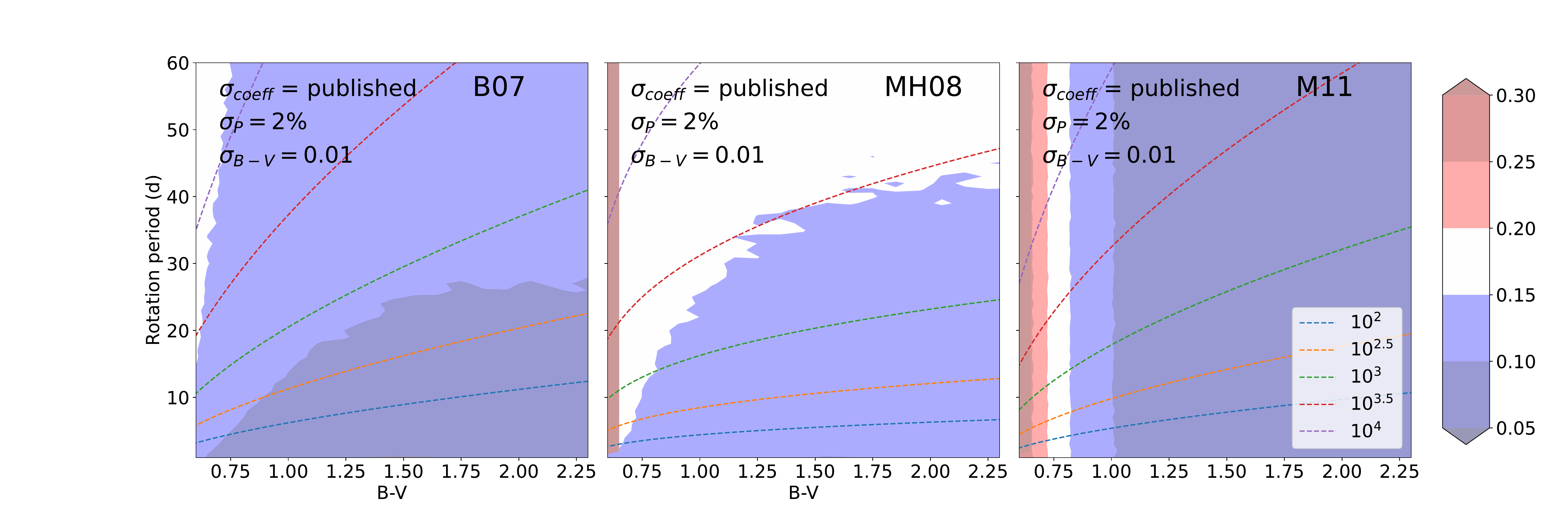}
\includegraphics[width=0.66\textwidth]{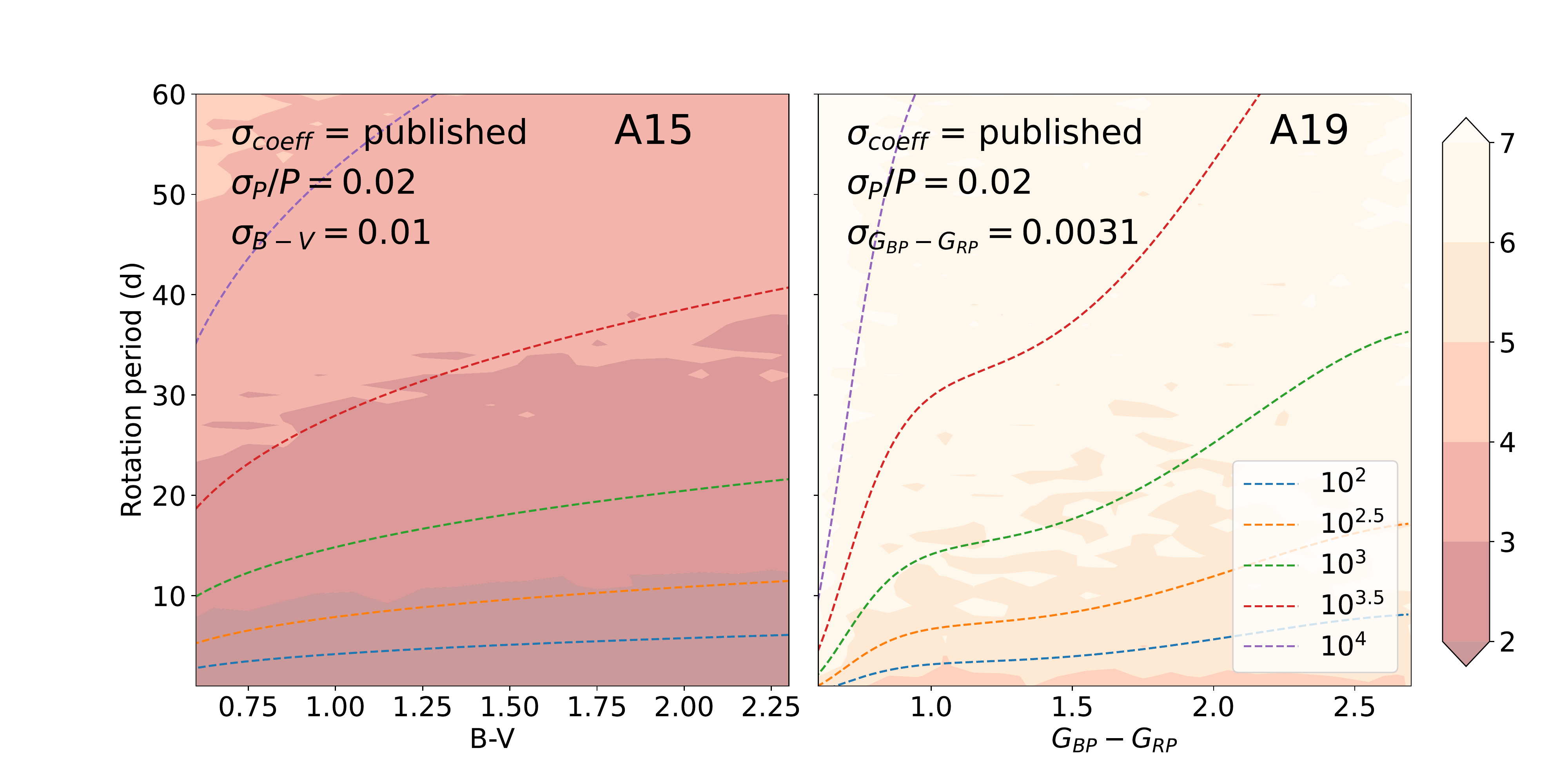}
\caption{Age uncertainty, $\frac{\sigma_{age}}{age}$, as a function of stellar color, period, and their uncertainties. Top: The age uncertainty for B07, MH08, and M11 models. Bottom: The age uncertainty for A15 and A19 models. For A19, only the gyrochronology component of the model was used. Note that the scale of A15 and A19 models (bottom) are different from the scale of the other models (top).  For the model uncertainties, we used the published values in Table~\ref{tab:coefficients} and \ref{tab:coefficients2}.  For the observational parameter uncertainties we used $\frac{\sigma_P}{P} = 2\%$ and  $\sigma_{B-V} = 0.01$.  We set $\sigma_{G_{BP}-G_{RP}} = 0.033$ based on the $G$ mag of those targets \citep{Evans2018}. \label{fig:contour_P}}
\end{center}
\end{figure}

Figure~\ref{fig:contour_P} presents the ranges of expected age precision in the form of this normalized age uncertainty, $\frac{\sigma_{age}}{age}$, for the four gyrochronology models of B07, MH08, M11, and A15.  We emphasize that these are age precision values, not age accuracy values, because they do not incorporate any systematic errors in the models. The fractional rotation period uncertainties, $\frac{\sigma_P}{P}$, for the primary and secondary were both set at 2\% of the observed period, which is the smallest uncertainties among the wide binary and triple star examples that B07 published (36 Oph A, B, and C). The $B-V$ uncertainties of 0.01, which B07 used in his analysis, was also incorporated.  For uncertainties of $G_{BP}$ - $G_{RP}$, we set $\sigma_{G_{BP}-G_{RP}} = 0.033$ based on the $G$ mag of those targets \citep{Evans2018}. For the coefficient uncertainties, the published values listed in Table~\ref{tab:coefficients} were used.  The uncertainties in model coefficients, observed rotation period, and colors were propagated via this Monte Carlo approach to derive uncertainties in the estimated ages. 

Figure~\ref{fig:contour_P} shows how age uncertainties depend on the uncertainties in model parameters, color, and rotation period.  For B07, MH08, and M11 models, stars with $\sigma _{P}$/$P$ = 0.10, the resulting age uncertainties are about 5-20\% for stars with $B-V \geq 1$ and above 20\% for stars with $B-V \leq 1$. B07 used an error propagation method to calculate the uncertainties of his model and found that they are 20\% for  stars with $B-V$ = 0.5, 15\% for stars with $B-V$ = 0.65, 13\% for the stars with $B-V$ = 1.0 or 1.5, and always greater than 11\%. Our results are similar to his, however our results indicate slightly larger age uncertainties (15-20\%) for stars with $B-V$ = 0.65 and slightly smaller uncertainties (5-10\%) for stars with $B-V$ = 1.5 and $P$ $\leq$ 20 days.  Although these age uncertainties may be too large as the basis for some research projects, e.g. refining the formation history of the Milky Way, they may be sufficient for other projects such as interpreting the radiation environment of an exoplanet. Using these models with high quality rotation rates ($\sigma _{P}$/P = 0.05) is necessary within these models to keep age uncertainties under 20\%.

The coefficient uncertainties of the A15 and A19 models are significantly larger than the coefficient uncertainties of the other two models, and these models therefore yield larger age uncertainties. This does not mean that the A15 and A19 models are of lower quality. Rather, they might simply be incorporating more realistic uncertainties.\footnote{One of the reasons that A15 model obtained large coefficient uncertainties is that the model included old asteroseismic stars that have stopped spinning down \citep{vanSaders2016}. They suggest that these large uncertainties only should apply to stars with large Rossby numbers, $Ro$ = $P$/$\tau_{cz}$.}
We have chosen these five models because their analytical form allows us to develop and demonstrate our Monte Carlo approach to testing gyrochronology models.

\section{Model Evaluation Method using Wide Binaries}  \label{sec:method}

As outlined in Section~\ref{sec:intro}, we can use wide binaries to evaluate gyrochronology models. We assume the two components are coeval \citep{Kraus2009}, ratio the two gyrochronological ages, and expect this ratio to be close to one.  In this section, we propose a method to assess four Skumanich-based (B07, MH08, M11, A15) and two non Skumanich-based (A19 and SL20) models.

For Skumanich-based models, assuming Equation~\ref{eq:1} applies, the rotation periods of the primary and secondary components of a binary system are 

\begin{equation}
P_p = (A_{g,p})^n \times a[(B-V)_p - c]^b
\label{eq:7}
\end{equation}

\noindent and 

\begin{equation}
P_s = (A_{g,s})^n \times a[(B-V)_s - c]^b
\label{eq:8}
\end{equation}

\noindent where a, b, c, and n are the parameters listed in Table~\ref{tab:coefficients}, $P_p$ and $P_s$ are the primary and secondary rotation periods, and $[B-V]_p$ and $[B-V]_s$ are the primary and secondary $B-V$ colors.

Although the model parameters (a, b, c, n) have uncertainties, for any given value of those parameters the ages of the two stars must be the same (e.g. $A_p = A_s$).  Taking the ratio of equations~\ref{eq:3} and \ref{eq:4} yields 

\begin{equation}
\frac{P_p}{P_s} = \left(\frac{A_{g,p}}{A_{g,s}}\right)^n \: \left[\frac{(B-V)_p - c}{(B-V)_s - c}\right]^b.
\label{eq:9}
\end{equation}

Solving for the age ratio yields 

\begin{equation}
\frac{A_{g,s}}{A_{g,p}} = \left\{\left[\frac{(B-V)_p - c}{(B-V)_s - c}\right]^b/\left(\frac{P_p}{P_s}\right)\right\}^{1/n}, 
\label{eq:age_comp}
\end{equation}

\noindent which reduces the number of parameters, since the coefficient $a$ drops out of the ratio.  Therefore, we do not test the absolute value of the age parameter, only the consistency of the ages derived for the two components.  Nominally, we expect $\frac{A_{g,s}}{A_{g,p}}$ = 1. 

We incorporated an additional parameter, $\Delta$, as an age ratio ``tolerance''.  This parameter represents the degree to which a gyrochronology model is imperfect and yet still tolerably useful.   Our goal is to use this age tolerance parameter to explore the required precision of the models and observational parameters in order to achieve useful gyrochronology ages.  For the remainder of this paper, we adopt $\Delta = 0.1$, i.e. a 10\% age difference between the primary and secondary can be tolerated and still deemed consistent.  Other values of $\Delta$ could be chosen, based on the age precision required for a particular project.  



We again assumed Gaussian or other distributions in the observed quantities and model parameters, as described above. Equation~\ref{eq:age_comp} is recast as

\begin{equation}
D\left(\frac{A_{g,s}}{A_{g,p}}\right) = \left\{\left[\frac{G(B-V)_p - D(c)}{G(B-V)_s - D(c)}\right]^{D(b)}/\left[\frac{G(P_p)}{G(P_s)}\right]\right\}^{1/D(n)}, 
\label{eq:11}
\end{equation}

\noindent where $G(B-V)$ and $G(P)$ are Gaussian distributions representing the observed colors and periods, respectively, and their uncertainties; $D(b)$, $D(c)$, and $D(n)$ are non-Gaussian distributions representing these model parameter values and their uncertainties.

For non Skumanich-based models, we adopted a similar approach.  We obtained the following equation using $G(B-V)_P$, $G(B-V)_S$ (or $Gaia$-base colors $G(G_{BP}-G_{RP})_P$, $G(G_{BP}-G_{RP})_S$ for A19), $G(P_P)$, and $G(P_S)$;

\begin{equation}
D\left(\frac{A_{g,s}}{A_{g,p}}\right) = \frac{f(G(B-V)_P, G(P_P))}{f(G(B-V)_S, G(P_S))}
\label{eq:non-S}
\end{equation}

\noindent where $f(G(B-V), G(P))$ is the function to obtain ages for the A19 and SL20 models.

\begin{figure}[ht!]
\plotone{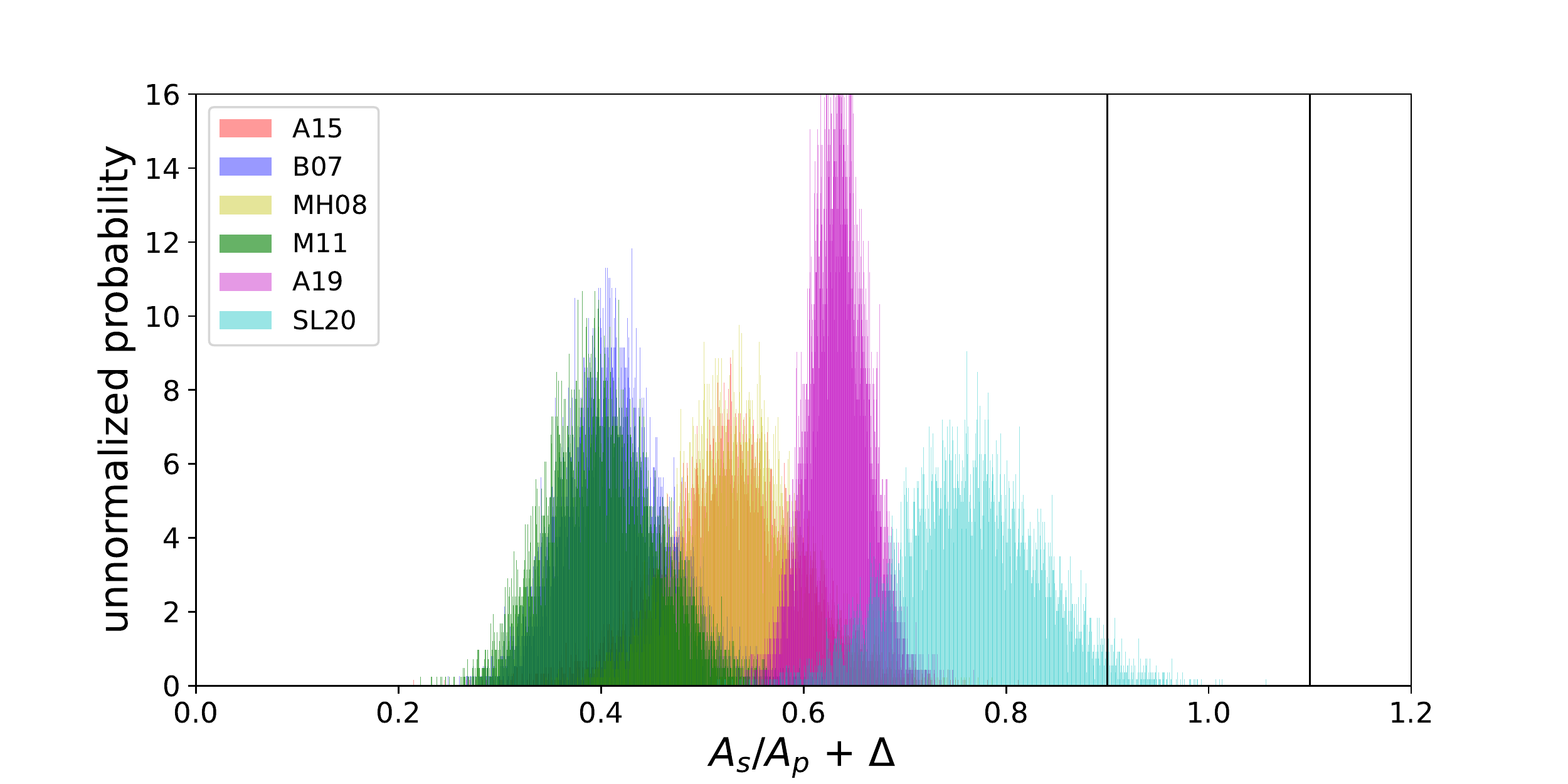} 
\caption{An example of the age-comparison probability distribution from Equation~\ref{eq:7}, using the models of B07, MH08, M11, A15, A19, SL20. HD 155885 and HD156026 are used for this example with the published periods and $B-V$ colors and their uncertainties (See in Table~\ref{tab:used_binary}). The vertical bar shows the region within 1 $\pm \Delta$, where $\Delta$ = 0.1.}
\label{fig:age-comparison-example}
\end{figure}

Figure~\ref{fig:age-comparison-example} displays examples of Equation~\ref{eq:7} \& \ref{eq:non-S} via the age-ratio probability distribution using each model for the color and period values for HD 155885 and HD 156026, two components of a triple star system listed in \citet{Barnes2007}. In this figure the models parameters were drawn randomly from their quoted error distributions and applied to Equation~\ref{eq:7} to calculate $\frac{A_{g,s}}{A_{g,p}}$.  This process was repeated 10,000 times in order to derive the age ratio distribution for this binary.  

The rotation periods and colors of the stars displayed in Figure~\ref{fig:age-comparison-example} are listed in Table~\ref{tab:used_binary}.  For this binary, the age ratio probability distributions of all of these models are inconsistent with $\frac{A_{g,s}}{A_{g,p}}$ = 1 $\pm$ $\Delta$ = 0.90 to 1.10.  The estimated secondary star age is younger than the primary age for all six models to a degree that exceeds uncertainties in the observational data or model parameters. 

To evaluate the degree to which the primary and secondary ages for any binary are consistent within the tolerance $\Delta$, the fractions of the probability distributions within 1 $\pm$ $\Delta$ were computed for each star, ratioed, and converted to a percentage as follows:

\begin{equation}
{\rm Coeval~probability} = 100 \times \left(\frac{\displaystyle\int_{1 - \Delta = 0.9}^{1 + \Delta = 1.1} \frac{A_{g,s}}{A_{g,p}}}{\displaystyle\int_{0}^{\infty} \frac{A_{g,s}}{A_{g,p}}}\right)
\label{eq:13}
\end{equation}

This value corresponds to the likelihood that the binary pair is coeval within the adopted tolerance (here 10\%) and assumed model.  Larger values from this equation increase confidence in the consistency of the model and the veracity of the observed binary properties.  If this value is low, it is likely that the stellar properties were not measured adequately or that there is a problem with the gyrochronology model itself.  Follow-up observations can determine whether the former is the case.  If not, adjustment to the empirical gyrochronology model may be called for. We adopted 99.7\% ($\geq$ 3 $\sigma$) as the threshold for discriminating binary components with discrepant ages. In the following section, we use the same formalism obtained above to examine a related issue -- the age precision possible with current and hypothetically improved uncertainties in the empirical model constants and the observed parameters.  

\section{Examples and Discussion} \label{sec:result}

\begin{figure}[ht!] 
\begin{center}
\includegraphics[width=\textwidth]{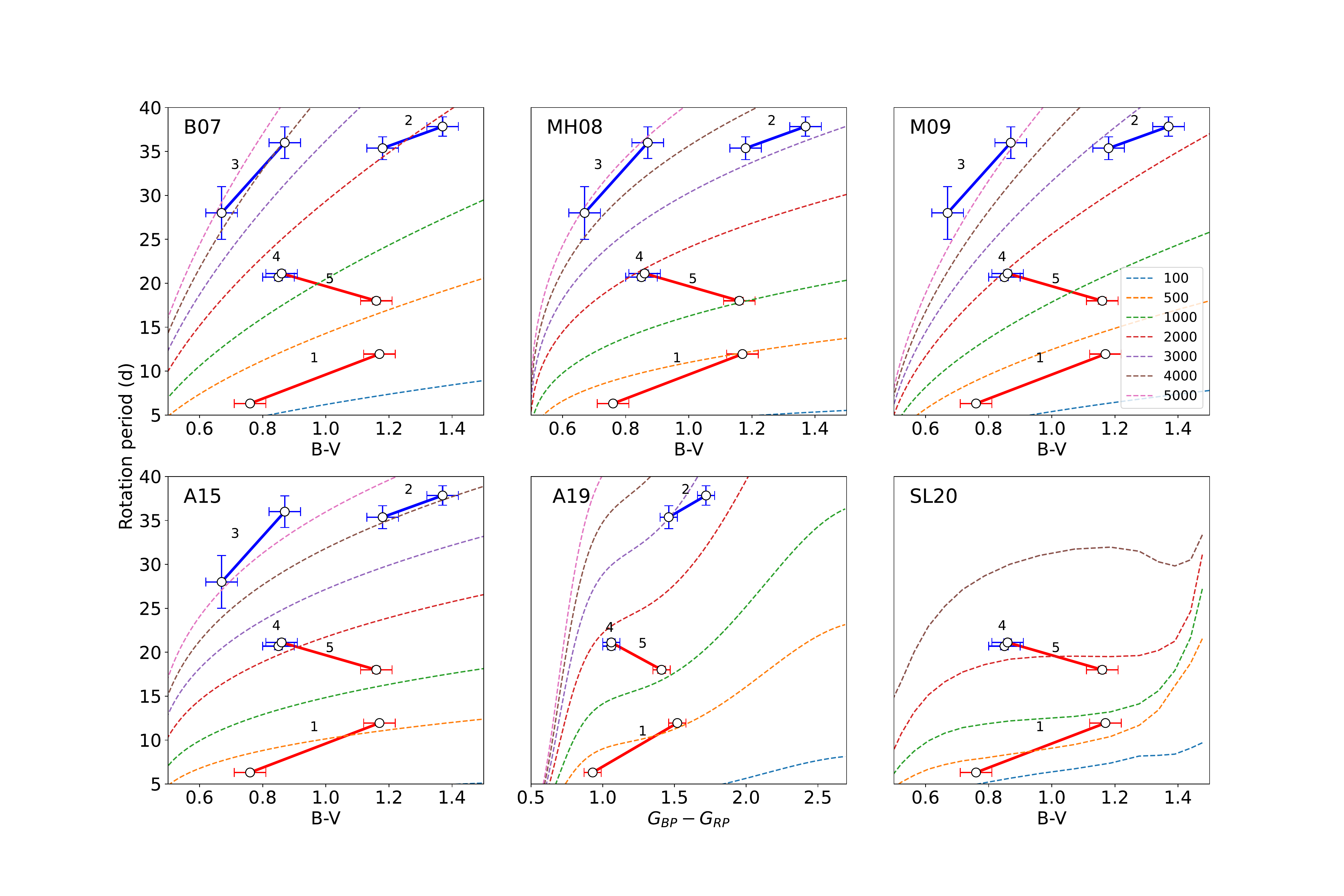}
\caption{Color-period diagrams for example wide binaries (1, 2, 3) and a triple system (4, 5) in the model grids of B07 (upper left), MH08 (upper middle), M09 (upper right), A15 (lower left), A19 (lower middle), and SL 20 models (lower right). The age of each gyrochrone (in Myr) is listed in the legend of the upper right panel. The crosses indicate the observed color and rotation period of each target along with their uncertainties. Because the components are coeval, the line connecting the two components of a binary should lie on a gyrochrone.  Blue symbols and lines indicate that both components have consistent ages for these gyrochronology models.  Red symbols and lines indicate that both components have inconsistent ages.  The A19 panel does not include wide binary 3 because $Gaia$ photometry for these targets is not available.  The SL20 panel does not include binaries 2 or 3 because this model is not available in the age range of those stars. \label{fig:one_example_gyro}}
\end{center}
\end{figure}

The wide binaries discussed in B07 were selected to illustrate the model evaluation analysis described in Section~\ref{sec:method}.  Among theses samples, B07 chose pairs that had both rotation period and B-V information.  He also used the systems without rotational interaction between the components (except 36 Oph A and B)  The colors and rotation periods of each component are listed in Table~\ref{tab:used_binary}.  The dereddened $G_{BP}-G_{RP}$ was calculated using bolometric corrections to  $G_{BP}$ and $G_{RP}$ \citep{Casagrande2018}.  The rotation periods and colors of these wide binary stars are shown in the color-period diagrams in Figure~\ref{fig:one_example_gyro} along with the gyrochrone grids of the B07, MH08, M11, A15, A19, and SL20. As examples, we plot three cases (blue pairs in Figure~\ref{fig:one_example_gyro}) where the ages of the two components appear to be consistent using all three gyrochronology models and two cases (red pairs in Figure~\ref{fig:one_example_gyro}) in which the ages of the two components appear to be inconsistent.  


The lines connecting the components of age-consistent pairs are nearly parallel to a gyrochrone while those of age-inconsistent pairs are clearly not parallel to any gyrochrone.  An eyeball assessment in this diagram provides a helpful age consistency check, though to determine quantitative age consistencies on must incorporate the observed and model uncertainties, the latter of which are not apparent in this diagram.  

\begin{deluxetable}{llllll} [ht!]
\tablecaption{Wide binaries and a triple star system} used in the experiment presented in Figure set~\ref{fig:BV_min_P_diff1}. The rotation period, rotation period uncertainty, and de-reddened (B-V) are obtained from B07. 1, 2, and 3 are binary systems and 4, and 5 are in the same triple star system.   Components of 3 do not have Gaia $G_{BP} - G_{RP}$ because those apparent magnitudes are brighter than  the Gaia bright limit.  \label{tab:used_binary}
\tablenum{3}
\tablehead{\colhead{internal ID} & \colhead{name} & \colhead{period} & \colhead{$(B-V)_0$} & \colhead{$(G_{BP}-G_{RP})_0$} & \colhead{type}\\
\colhead{} & \colhead{} & \colhead{(days)} & \colhead{(mag)} & \colhead{(mag)}& \colhead{}} 
\startdata
1 & $\xi$Boo A (HD131156 A)    & 6.31 (0.05) & 0.76 (0.01) & 0.93 (0.06)& G8V\\
         & $\xi$Boo B (HD131156 B)     & 11.94 (0.22) & 1.17 (0.01) & 1.52 (0.06)& K4V\\
2 & 61 Cyg A (HD 201091) & 35.37 (1.3) & 1.18 (0.01) & 1.46 (0.06)& K5V\\
         & 61 Cyg B (HD 201092) & 37.84 (1.1) & 1.37 (0.01) & 1.72 (0.06)& K7V\\
3 & $\alpha$Cen A (HD 128620) & 28 (3.0) & 0.67 (0.01) & N/A& G2V\\
         & $\alpha$Cen B (HD 128621) & 36 (1.8) & 0.87 (0.01) & N/A& K1V\\
4 & 36 Oph A (HD 155886) & 20.69 (0.4) & 0.85 (0.01) & 1.06 (0.06)& K1\\
         & 36 Ohp B (HD 155885) & 21.11 (0.4) & 0.86 (0.01) & 1.06 (0.06)& K1\\
5 & 36 Oph B (HD 155885) & 21.11 (0.4) & 0.86 (0.01) & 1.06 (0.06)& K1\\
         & 36 Oph C (HD 156026) & 18.0 (0.4) & 1.16 (0.01) & 1.41 (0.06)& K5\\
\enddata
\end{deluxetable}

\clearpage

\begin{figure}[ht!] 
\plotone{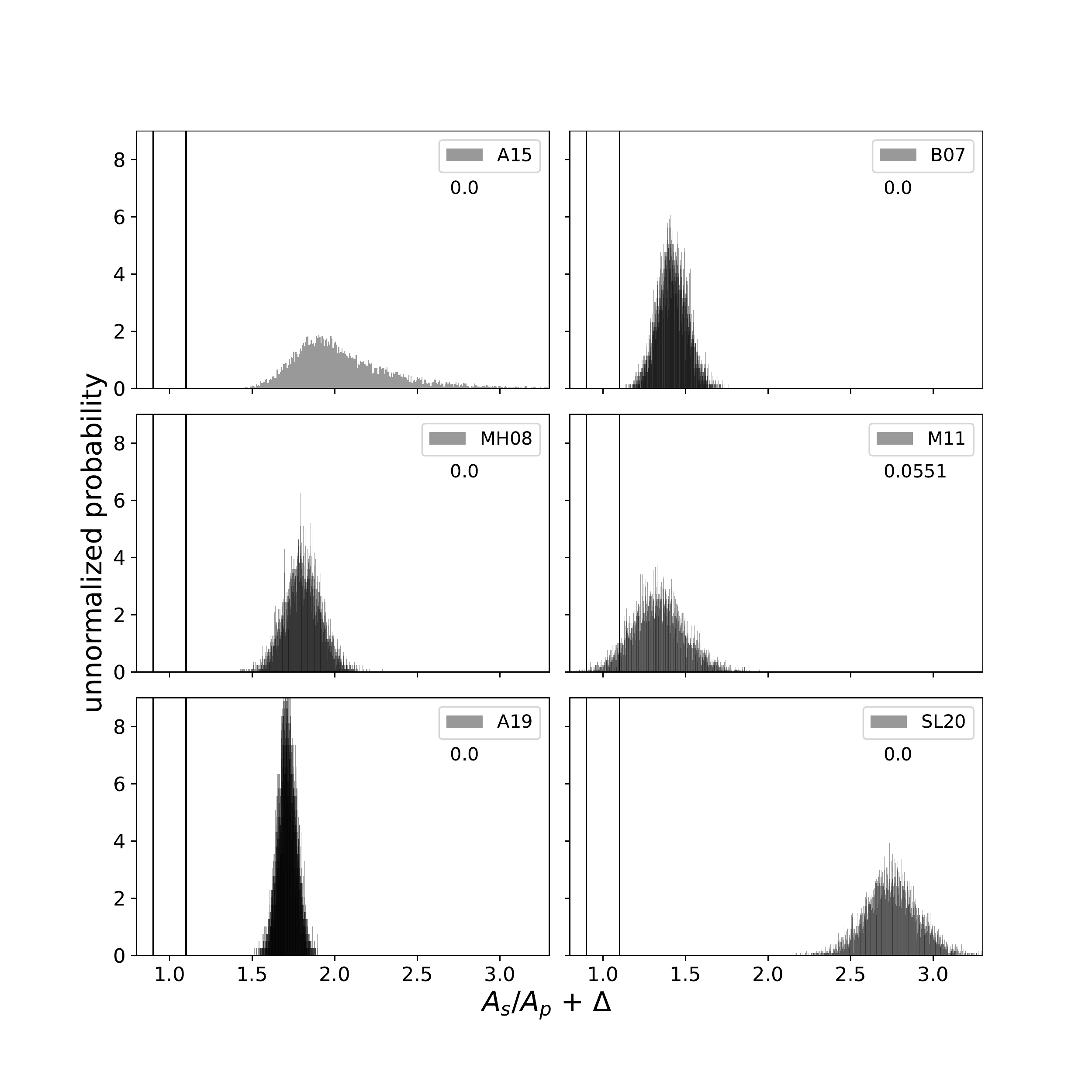} 
\caption{Comparison of age ratio probability distributions of components of binary 1 ($\xi$Boo A/B) whose connecting line does not parallel a gyrochrone which does not parallel a gyrochrone in Figure~\ref{fig:one_example_gyro}.  Each gyrochronology model (A15, B07, MH08, M11, A19, \& SL20) is used for each subplot. The percentage of each distribution that is within the tolerance area is shown in parentheses in the lower right in each panel. A ten percent tolerance ($\Delta$ = $\pm$0.10) in age agreement between components of such binary is illustrated by the vertical black lines. Comparison of age ratio probability distributions of components of each binary in shown in the Figure set (5 images), which is available in the online journal.\label{fig:BV_min_P_diff1}}
\end{figure}

Figure set~\ref{fig:BV_min_P_diff1} displays the corresponding age ratio probability distributions of the components of these binary and triple star systems. As listed in Table~\ref{tab:used_binary}, all targets are main sequence stars and all targets except $\xi$Boo A and $\alpha$Cen A are K dwarfs.  For each model, the fraction of the age ratio distribution where the two components are consistent within the tolerance level ($\Delta$ = $\pm$0.1), is listed in Table~\ref{tab:ratio}.

\begin{deluxetable}{lllllll} [ht!]
\tablecaption{Binary component consistency fractions for five pair-wise comparisons and six models.} \label{tab:ratio}
\tablenum{4}
\tablehead{\colhead{components} & \colhead{B07} & \colhead{MH08} & \colhead{M11} & \colhead{A15}& \colhead{A19} & \colhead{SL20}} 
\startdata
$\xi$Boo A/B    & 0.0000 & 0.0000 & 0.0579 & 0.0000 & 0.0000 & 0.0000 \\
61 Cyg A/B      & 0.4194 & 0.7389 & 0.4139 & 0.6991 & 0.3795 & N/A    \\
$\alpha$Cen A/B & 0.2741 & 0.3392 & 0.2036 & 0.2789 & N/A    & N/A    \\
36 Oph A/B      & 0.8388 & 0.8798 & 0.8309 & 0.8435 & 0.9064 & 0.9251 \\
36 Oph B/C      & 0.0000 & 0.0000 & 0.0000 & 0.0000 & 0.0000 & 0.0271 \\
\enddata
\end{deluxetable}

Binary no. 1 ($\xi$Boo A/B) is a wide binary consisting of components with spectral types G8V and K4V. B07 obtained an age of 187 Myr and 265 Myr for $\xi$Boo A and B, respectively. Our result eliminates the possibility that the ages of the two component are consistent within the tolerance ($\Delta$ = $\pm$0.1) for any models except M11.  One of the possible reasons for these inconsistent ages is that the stars are too young for the Skumanich law given subsequent work and the measurement that $\xi$Boo A (P = 6.31 days) may not yet be on the show rotator sequence.

Binary no. 2 (61 Cyg A/B) consists of a K5V+K7V wide binary pair.  The estimated ages by B07 are 2.12 and 1.87 Gyr, respectively.  The probability that the component ages are consistent within the tolerance is larger for A15 and MH08 models than other models.

Binary no. 3 ($\alpha$Cen A/B) consists of a G2V+K1V pair.  The derived ages by B07 are 4.6 and 4.1 Gyr, respectively.  The component ages appear to be consistent, though that may only be because the age ratio distributions are so wide, which in turn is due to the large  rotation period uncertainty of $\alpha$Cen A.

Triple star system (no. 4 and 5) consists of two chromospherically active K1 dwarfs (36 Oph A/B) and a K5 dwarf at much greater separation (36 Oph C).  B07 warns that A and B could potentially have interacted rotationally.  The ages he determined for A and B are both 1.43 Gyr, but the age of C is 590 $\pm$ 70 Myr.  He mentioned the possibility that A and B may have interacted and thereby decreased their period.  Our result agrees with this age discrepancy between A/B and C. A and B ages are consistent for all the models (0.83 to 0.93 of the entire distribution are within the tolerance level for all models).  However, B and C ages are inconsistent for all models (0 to 0.03 of the entire distribution are within the tolerance level for all models.

The color uncertainties are small for these example systems.  The rotation period uncertainties are also small for the majority of them.  However, our age comparison method does not work well (nor would any other method work well) if either rotation period or color uncertainties are large.  Figure set~\ref{fig:BV_min_P_diff1} demonstrates this in detail.  Since the  period uncertainties for $\alpha$Cen A are large ($\sigma_P/P$ = 0.11), these  distributions are broader than most of the other distributions. Thus, even though it may appear that the ages of the two components are consistent in the sense that they overlap 1.0, there is actually a greater probability that the ages are inconsistent, i.e. outside the 1 $\pm~\Delta$ range, than consistent and within 1 $\pm~\Delta$. Most of the original $Kepler$ periods have reported uncertainties around $\sigma_{P}/P$ = 0.05 \citep{Aigrain2015}. However, for $K2$, typical uncertainties are larger, perhaps because the observation span is much shorter than the original $Kepler$ data (see Fig. 4 of \citet{Reinhold2020}). Therefore, K2 data may be insufficient for precise enough periods, particularly for slower rotators. The best $B-V$ uncertainties that can generally be obtained from ground-based telescopes are $\sim$~0.02 mag \citep{Tonry2012}, however most of the $TESS$ Input Catalog \citep{Stassun2018} entries have $\sigma_{B-V} \geq$ 0.05. On the other hand, colors obtained from Pan-STARRS and  $Gaia$ often have substantially smaller uncertainties. Pan-STARRS $g-r$ uncertainties are around 0.04 and $Gaia$ $G_{BP}-G_{RP}$ uncertainties are 0.002 at $G$ $<$ 13, and 0.01 at $G$ = 17 \citep{Tonry2012, Evans2018}. Therefore, we suggest that the next generation of gyrochronology relations be based on photometry from $Gaia$ and Pan-STARRS, or other large-field, high-quality surveys, for example as done by A19.   

\clearpage

\figsetstart
\figsetnum{4}
\figsettitle{Comparison of age ratio probability distributions of components of binaries}

\figsetgrpstart
\figsetgrpnum{4.1}
\figsetgrptitle{Comparison of age ratio probability distributions of components of binary 1 ($\xi$Boo A/B) whose connecting line does not parallel a gyrochrone which does not parallel a gyrochrone in Figure~\ref{fig:one_example_gyro}.  Each gyrochronology model (A15, B07, MH08, M11, A19, \& SL20) is used for each subplot. The percentage of each distribution that is within the tolerance area is shown in parentheses in the lower right in each panel. A ten percent tolerance ($\Delta$ = $\pm$0.10) in age agreement between components of such binary is illustrated by the vertical black lines.}
\figsetplot{HD131156A_B_distribution.pdf}
\figsetgrpend

\figsetgrpstart
\figsetgrpnum{4.2}
\figsetgrptitle{Similar to Figure 4.1, but for binary 2 (61 Cyg A/B), which approximately parallels a gyrochrone in Figure~\ref{fig:one_example_gyro} for some models.}
\figsetplot{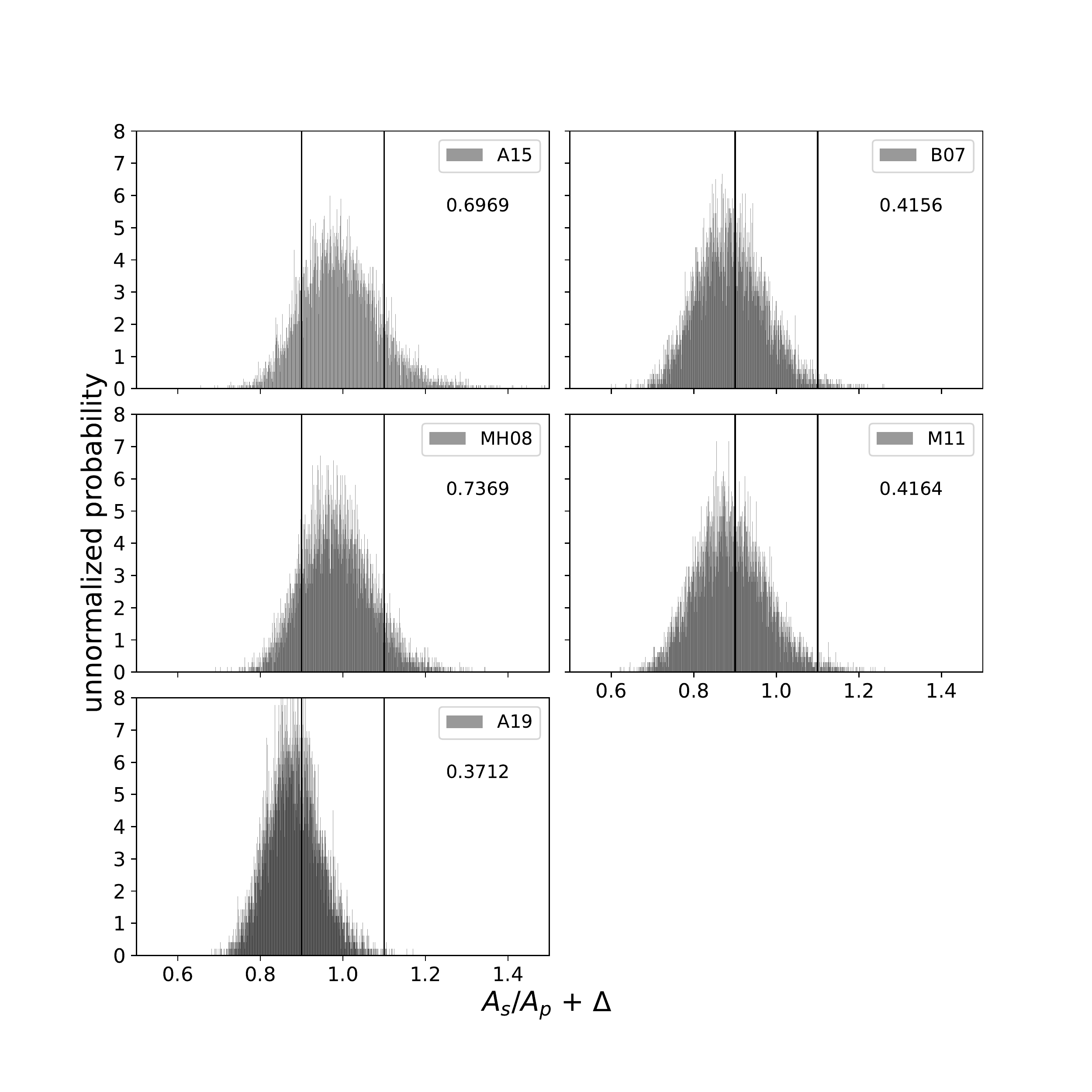}
\figsetgrpend

\figsetgrpstart
\figsetgrpnum{4.3}
\figsetgrptitle{Similar to Figure 4.1, but for binary 3 ($\alpha$Cen A/B), which approximately parallels a gyrochrone in Figure~\ref{fig:one_example_gyro}.  The A19 model is not shown here because the $G_{BP} - G_{RP}$ of these targets are not available. The SL20 model is not shown because the model was not available for the period and $B-V$ ranges of the targets.}
\figsetplot{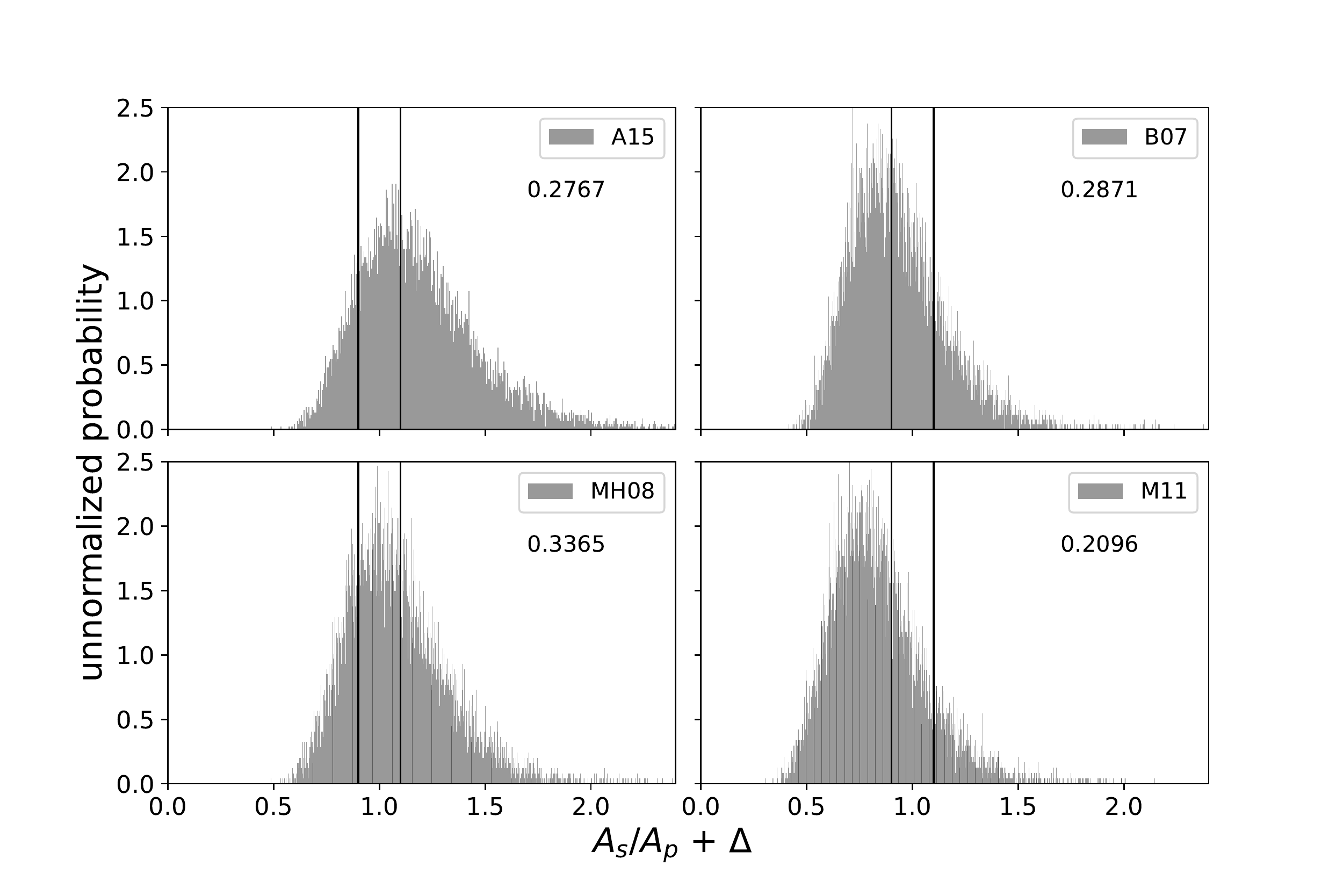}
\figsetgrpend

\figsetgrpstart
\figsetgrpnum{4.4}
\figsetgrptitle{Same as Figure 4.1, but for binary 4 (36 Oph A/B), which approximately parallels a gyrochrone in Figure~\ref{fig:one_example_gyro}.}
\figsetplot{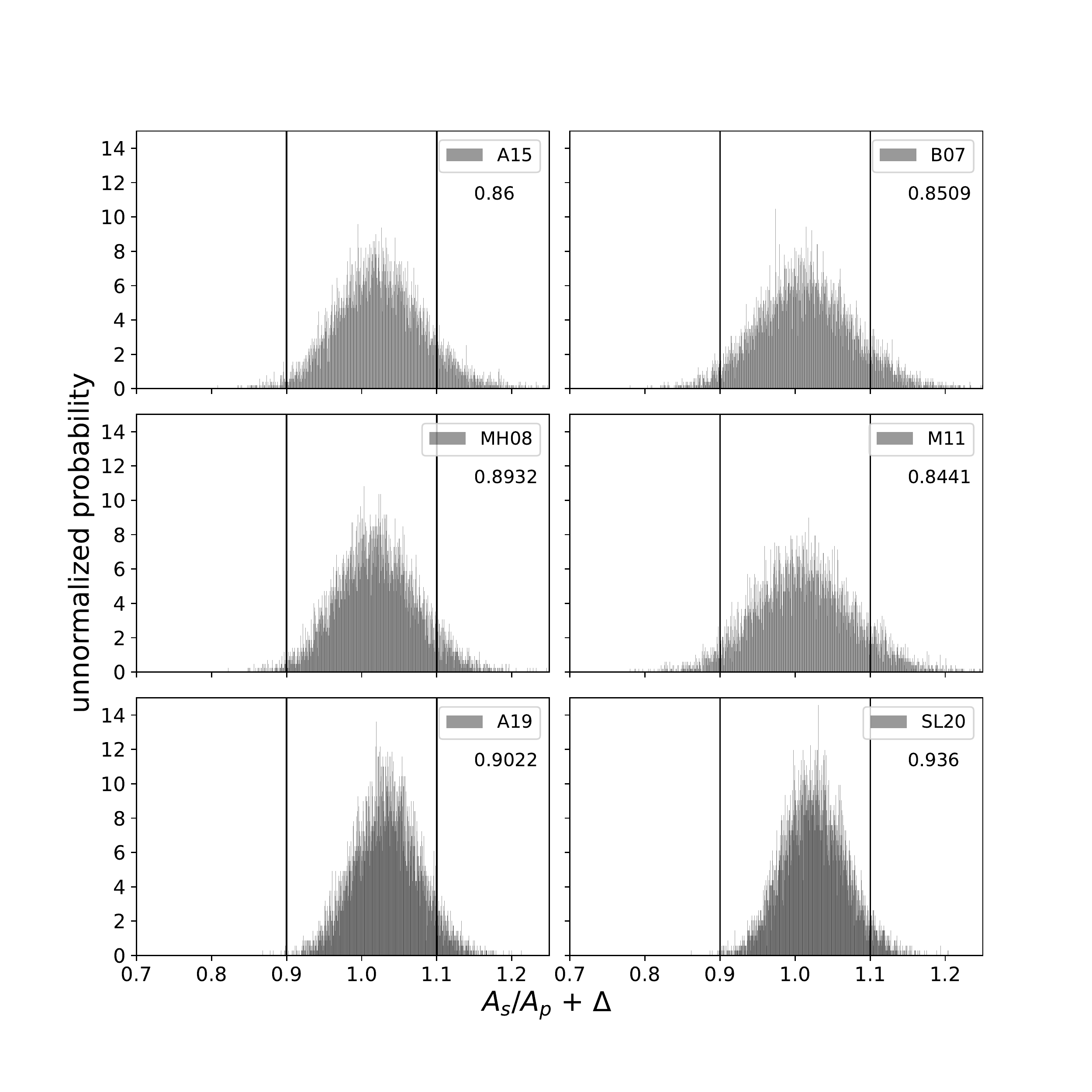}
\figsetgrpend

\figsetgrpstart
\figsetgrpnum{4.5}
\figsetgrptitle{Same as Figure 4.1, but for binary 5 (36 Oph B/C), which does not parallel a gyrochrone in Figure~\ref{fig:one_example_gyro}.}
\figsetplot{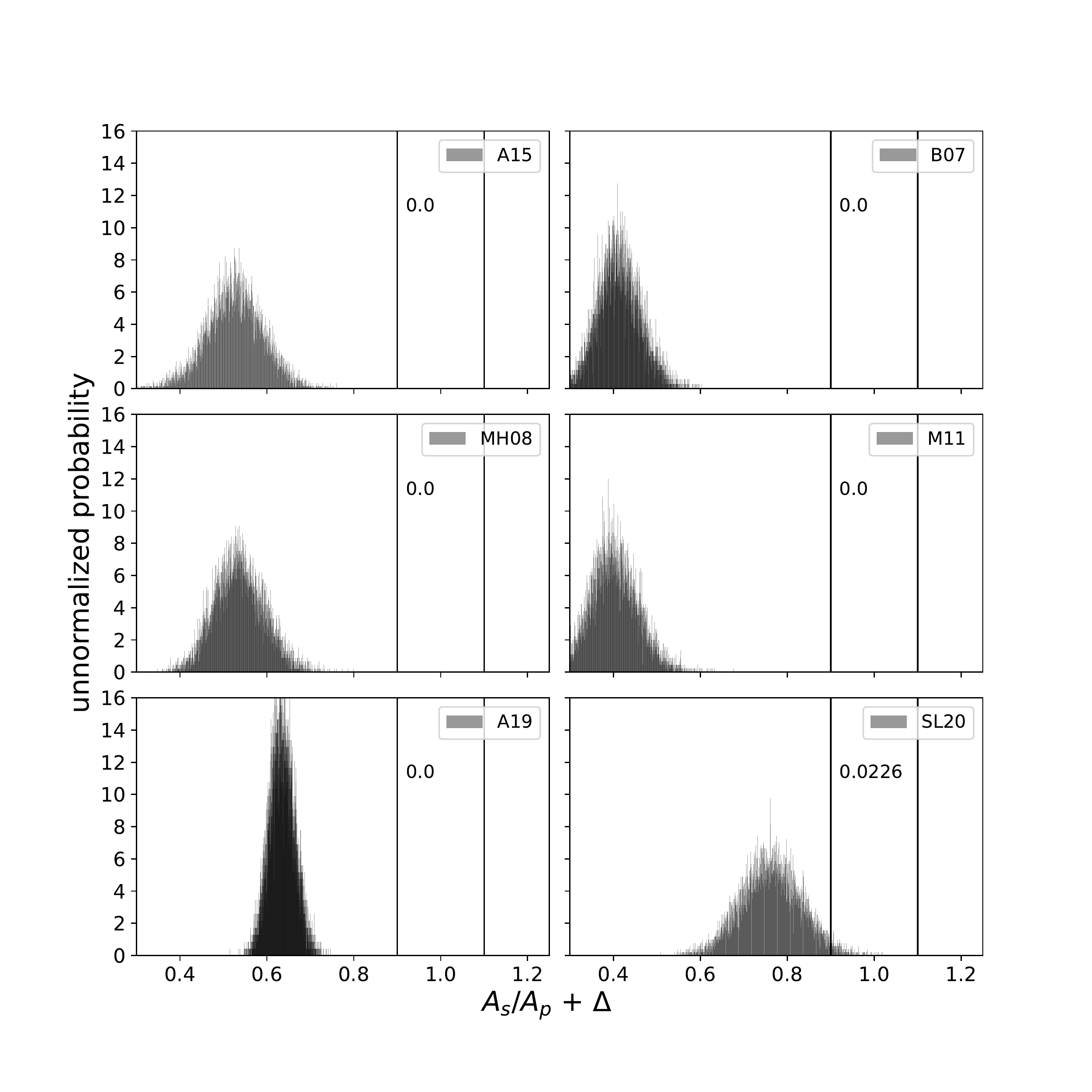}
\figsetgrpend

\figsetend

\section{Conclusion} \label{Conclusions}

Main sequence stars with masses of less than 1.3 $M_\sun$ lose their angular momentum as they age and thus their rotation periods increase over time. Although current gyrochronology concepts work relatively well for young MS stars with masses between 1.0 and 1.3 $M_\sun$, we have learned in the past ten years that improvements to the models are required for older and lower mass MS stars.
In this paper, we propose a numerical model evaluation method using wide binaries under the assumption that binary components are coeval. We test six models that rely only on color and period information (B07, MH08, M11, A15, A19, and SL21) as examples. The use of binary stars supplements open cluster data because the abundance of binaries allows us to cover and fill in a wider range of stellar age and metallicity.  We use our model evaluation method to determine the probability that binary components have consistent ages, finding examples where component stars are consistent as well as examples where ages are inconsistent.  Both consistent and inconsistent pairs can help the community improve data quality and model fitting, and ultimately the precision of gyrochronology ages.
In addition, we used a Monte Carlo approach to assess the precision and accuracy of ages derived from four Skumanich-based models (B07, MH08, M11, A15)).  The age uncertainties that we obtained are similar to those obtained by B07 and are  $\sigma_{age}$/$age$ = 5-20 \% for $\sigma_{P}$/$P$ = 2\% and $\sigma_{B-V}$ = 0.01.  Our results show that gyrochronological age uncertainty is slightly higher than derived by B07 for stars with $B-V$ = 0.65 ($\sigma_{age}$/$age$ = 15-20\%) and slightly smaller for the stars with $B-V$ = 1.5 and rotation period $P \leq$ 20 days ($\sigma_{age}$/$age$ = 5-10\%).  We also found that with the traditional Skumanich-based gyrochronology models of B07, MH08, M11, and A15, gyrochronology ages can be determined to approximately 5 to 20\% if $\sigma_{B-V} \leq$ 0.01 and $\frac{\sigma_P}{P} \leq$ 0.02, that are the smallest uncertainties among binary samples of B07 (Please note that this is the precision claimed by these specific models and this is not the actual precision).

\acknowledgements

An anonymous referee has provided valuable insight through questions posed in the report on the draft of this paper, and for which the authors are most grateful. \\

This material is based upon work supported by the National Science Foundation under Grant No.\ AST-1715718, \ AST-1910396, and\ AST-2108975 and by NASA under Grant No. NNX16AB76G and 80NSSC21K0245. \\

\end{document}